\documentclass[12pt, usenatbib]{article}

\usepackage{amssymb}
\usepackage[authoryear]{natbib}
\usepackage{bm}
\usepackage{indentfirst}
\usepackage{graphics}
\usepackage{amsmath}
\usepackage{epsfig}
\usepackage{threeparttable}
\usepackage{amsthm}
\usepackage{latexsym}
\usepackage[figuresright]{rotating} 
\allowdisplaybreaks

\usepackage{enumitem}

\newfont{\Sc}{eusm10}

    \def\independenT#1#2{\mathrel{\setbox0\hbox{$#1#2$}%
    \copy0\kern-\wd0\mkern4mu\box0}}

\usepackage[font={small}]{caption}
\usepackage{setspace}
\begin{document}

\begin{center}
\large 

\vspace{1in}

Probability-Scale Residuals in HIV/AIDS Research: Diagnostics and Inference

\normalsize
\bigskip

\bigskip
\bigskip
\bigskip
\bigskip
\bigskip

Bryan E. Shepherd$^1$,  Qi Liu$^2$, Valentine Wanga$^3$, Chun Li$^4$

\bigskip
\bigskip
\bigskip
\bigskip
\bigskip
\bigskip
\bigskip
\bigskip

%\begin{align*}
%            ^1      &  \text{Department of Biostatistics} \\
%                  &  \text{Vanderbilt University School of Medicine}  \\
%                  &  \text{2525 West End Suite 11000}  \\
%                  &  \text{ Nashville, Tennessee  37203, U.S.A.} \\
%                  &  \text{Phone: 615-343-3496} \\
%                  &  \text{Fax:  615-343-4924} \\
%                  &  \text{bryan.shepherd@vanderbilt.edu} \\
%                  &  \text{qi.liu.1@vanderbilt.edu} \\ \\ \\
%            ^2   &  \text{Department of Epidemiology and Biostatistics} \\
%                  &  \text{Case Western Reserve University}  \\
%                  &  \text{Cleveland, Ohio 44106, U.S.A.} \\
%                  &  \text{cxl791@case.edu}
%\end{align*}   

$^1$ Department of Biostatistics, Vanderbilt University School of Medicine, \\ 2525 West End, Suite 11000, Nashville, Tennessee 37203  \\ email: bryan.shepherd@vanderbilt.edu

\bigskip
\bigskip

$^2$ Late Development Statistics, \\Merck \& Co., 126 E. Lincoln Avenue, Rahway, New Jersey 07065 \\ email: qi.liu4@merck.com

\bigskip
\bigskip

$^3$ Departments of Epidemiology and Global Health, University of Washington, \\ 1959 NE Pacific Street, Health Sciences Building, F-262, Seattle, Washington 98195 \\ email: wangav@uw.edu
%$^3$ Institute for Health Metrics and Evaluation, \\ 2301 Fifth Avenue, Suite 600, Seattle, Washington 98121 \\ email: wangav@uw.edu

\bigskip
\bigskip

$^4$ Department of Epidemiology and Biostatistics, Case Western Reserve University, \\ Wolstein Research Building 2528, Cleveland, Ohio 44106  \\ email:  cxl791@case.edu

\bigskip
\bigskip
\bigskip
\bigskip
\bigskip
\bigskip
\bigskip
\bigskip

\bigskip
\bigskip

%Short Running Title:  Probability-Scale Residuals

%\begin{align*}
%\text{Corresponding Author: } &  \text{ Bryan Shepherd} \\
%          &  \text{ 1161 21st Avenue South, S2323 MCN}  \\
%          &  \text{ Nashville, TN 37232-2158}
%\end{align*}           

\bigskip
\bigskip

\end{center}

\setlength{\baselineskip}{14pt}

\newpage

\raggedright

\def\baselinestretch{2}\small\normalsize%   1.5

%\large \textbf{ABSTRACT}

\normalsize

\bigskip
\bigskip
\bigskip

%\noindent KEY WORDS: Diagnostics; Generalized linear model;  HIV; Quantile regression; Rank statistics; Survival analysis.

\newpage

%\setcounter{page}{1}
%\pagenumbering{num_style}

\setlength{\parindent}{0.5in} 
\setlength{\baselineskip}{12pt}

\bigskip 

\def\baselinestretch{1.7}\small\normalsize%   1.5

\section{Introduction}

There is a wide variety of data in HIV/AIDS research.  In clinical studies, common variables include CD4 cell count, HIV-1 RNA (viral load), demographics, WHO or CDC disease stage, and time-to-event variables such as time from ART initiation to an AIDS defining event, viral failure, or death.  Biomarker data are common in both clinical and basic studies of HIV; these may include markers of inflammation, pharmacokinetics,  drug use, or metabolism, and may be biomarkers commonly used in other disease settings (e.g., diabetes or hepatitis).  Genomic data, both human and viral, are also important.  

Of course, the characteristics of these and other variables used in HIV research are extremely diverse.  The distribution of some are fairly symmetric (e.g., age at ART initiation), somewhat skewed (e.g., CD4 count), or highly skewed (e.g., viral load). Many variables are left censored at detection limits (e.g., viral load and other biomarkers) or right censored due to finite follow-up (e.g., time to death).  Many are ordered categorical (e.g., stage of disease and single nucleotide polymorphisms).  Much of the statistical work in the analysis of HIV data involves finding proper models for these variables to assess associations, to predict outcomes, and hopefully in the end to improve patient and public health. Given the diversity of variables, a wide variety of statistical models are used.

There are benefits to having statistical methods that are robust and efficient across a wide variety of data types.  Such methods can be quickly applied with confidence in many different situations and may be useful as a first pass in big data settings (e.g., datasets with many variables of interest).  Such methods may also be useful in smaller analyses because they provide nice, simple summaries.  Spearman's rank correlation is one such example: its simplicity, validity, and utility across a wide variety of orderable variables makes it popular in practice.  

We have developed a new type of residual, the probability-scale residual (PSR), which is remarkably useful and well defined across a wide variety of variable types and models \citep{li2010test,li2012new,shepherd2016psr}.  As a residual it can be used for model diagnostics and for inference.  We have proposed to use the correlation of PSRs to adjust Spearman's rank correlation for covariates \citep{liu2016}.  The goal of this chapter is to introduce the PSR to HIV researchers, to describe a few of its important properties, and to demonstrate its utility across a diverse set of analyses with HIV data.  We first introduce the PSR and describe some of its properties (section 2).  We then illustrate its use for model diagnostics (section 3) and inference (section 4).  Several datasets that we have encountered in our collaborative HIV/AIDS research will be used to illustrate the methods.  The final section discusses a few points and proposes directions for future research.  Analysis code for all of the data examples is available at our website, biostat.mc.vanderbilt.edu/ArchivedAnalyses.

\section{Probability-scale Residual}

In linear regression, a residual is defined as $y-\hat y$, where $y$ is an observed
value and $\hat y$ is a fitted value, typically the estimated expectation of the outcome conditional on covariates.  This observed-minus-expected residual (OMER) is simple and has many desirable properties, but is not easily extendable to outcomes where conditional expectations are difficult to calculate or are not meaningful.  For example, for ordinal outcomes there is no natural definition of difference or conditional expectation unless scores are assigned to the ordered categories; for right censored outcomes with partially defined fitted distributions one may not be able to calculate the conditional expectation.  Furthermore, the OMER may be misleading with models where one is fitting a non-symmetric distribution to data.  

The OMER can be thought of as the expectation of the difference between the observed value, $y$, and a random variable, $Y^*$,
from the fitted distribution with cdf $F^*$: $E(y-Y^*) = y-\hat y$. Instead of using the difference to contrast $y$ and $Y^*$, one could use the sign function, specifically, $\text{sign}(y,Y^*)$, where sign$(a,b)$ is $-1, 0,$ and 1 for $a<b, a=b,$ and $a>b,$ respectively. The PSR is simply the expectation of this contrast,  
$r(y,F^*)=E\left\{\text{sign}(y,Y^*)\right\}$, which can be written in terms of probabilities as $P(Y^*<y)-P(Y^*>y)$ or equivalently $F^*(y-)+F^*(y)-1.$

A few benefits of the PSR are immediately apparent.  First, the sign function is more generally applicable than the difference, so by contrasting variables with the sign function we are able to define a residual for more types of outcomes.  Second, the PSR does not require estimation of conditional expectation but rather estimation of the conditional distribution itself.  Note that it is not necessary to estimate the entire distribution, but the distribution only needs to be estimated at the observed value, $y$.  Hence, the residual is flexible for a wide range of outcome variables (including ordered categorical) and models (including semiparametric), while still providing information on model fit. 

We originally introduced the PSR for use with ordered categorical variables \citep{li2010test,li2012new}.  In that setting, the residual has several nice properties including the following: 
\begin{enumerate}
\item The PSR captures order information without assigning arbitrary numbers to the ordinal categories.  
\item The PSR yields only one value per observation regardless of the number of categories of the ordinal variable. 
\item The PSR has expectation zero with a correctly specified model; i.e., for a random variable $Y$ with distribution $F$, $E\{r(Y,F^*)\}=0$ if $F^*=F$.
\end{enumerate}
In addition to the above, the PSR is the only residual that satisfies several natural properties for ordinal outcomes such as the branching property \citep{brockett1977characterization}, reversability, and monotonicity with respect to the observed value.  Details are in \cite{li2012new}.  For proportional odds models, the PSR sums to 0.  

The PSR is also well defined for other types of orderable data, including binary, count, continuous, and censored outcomes \citep{shepherd2016psr}.  In all cases, the PSR has expectation zero under correctly specified models.  With continuous data, the PSR equals zero at the median of the fitted distribution.  In addition, with continuous data the random variable $r(Y,F^*)$ is uniformly distributed between $-1$ and 1 if the fitted distribution is correctly specified, suggesting that with sufficiently large sample sizes, one can approximately assess model fit by comparing the distribution of PSRs with a uniform($-1,1)$ distribution.  With binary data, the PSR is simply the OMER, or the unscaled Pearson residual.  With right censored data, the PSR is a function of the observed minimum of the censoring and event times, $y$, and the indicator that an event occurred, $\delta$:  $r(y,F^*,\delta)=F^*(y)- \delta \left\{1-F^*(y-)\right\}$.  Given $\delta$, the PSR with time-to-event data is a one-to-one function of the martingale, deviance, and Cox-Snell residuals.  The PSR can also be written for current status data.  Details are in \cite{shepherd2016psr}, where we also demonstrated the calculation and the utility of the PSR in a variety of settings including normal linear models, least squared regression, exponential regression models, median/quantile regression, semi-parametric transformation models, Poisson and negative binomial regression, Cox regression, and the analysis of current status data.

Although novel, the PSR is related to other methods proposed in the statistical literature.  The PSR was part of a test statistic in genetic analysis of ordinal traits \citep{zhang2006detection}. The PSR is closely related to ridits \citep{bross1958use} and can be thought of as a linear transformation of an observed value's adjusted rank (see Section 4).  With continuous data, the PSR is simply a re-scaling of the probability integral transformation, which has been previously proposed for assessing goodness of fit \citep{pearson1938, davidjohnson1948} and as a component of a residual \citep{coxsnell1968, davisontsai1992, dunnsmyth1996}.  One of the strengths of the PSR is its unification of several of these concepts into a single residual.

\section{Model Diagnostics}

As a residual, the PSR can be useful for model diagnostics.  In this section, we introduce a few HIV datasets and demonstrate the residual's use in these settings.  

\subsection{Stage of Cervical Lesions: Proportional Odds Models}

Cervical specimens from 145 HIV-infected women in Zambia were examined using cytology and categorized into five ordered stages; 10 specimens were normal,  26 atypical squamous cells of undetermined significance (ASCUS), 35 low grade intraepithelial lesions, 49 high grade intraepithelial lesions, and 30 suspicious for cancer \citep{parham2006}.  Other data were collected from the women including age, CD4 count, education, and frequency of condom use (never, rarely, almost always, always). There is interest in modeling the association between these variables and stage of cervical lesions.  To this end, we fit proportional odds models with stage of cervical lesions as the outcome and various combinations of the other variables as predictors.  The left panel of Figure 1 shows a residual-by-predictor plot with the x-axis showing age and the y-axis showing PSRs from a proportional odds model with age and CD4 included as linear predictors; a lowess curve demonstrating the smoothed relationship between the residuals and age is also included.  When age is included in the proportional odds model as a linear variable,  there appears to be a quadratic relationship between the PSRs and age:  the model tends to over-predict severity of lesions at low and high ages. For example, a 23-year old woman in the dataset with a CD4 count of 309 cells/mm$^3$ had predicted probabilities of 0.10, 0.25, 0.27, 0.27, and 0.11 for cytology being normal, ASCUS, low, high, and cancerous, respectively.  This suggests that her observed cytology of ASCUS was less severe than predicted by the model -- resulting in a residual of $0.10-(0.27+0.27+0.11)=-0.55$ (left-most residual in the left panel).  If both linear and quadratic terms of age are included in the proportional odds model, the quadratic relationship between the residuals and age is no longer seen (right panel), suggesting a better model fit.  The observed ASCUS cytology for the 23-year old woman is now more consistent to what the model predicts (probability of normal, ASCUS, low, high, and cancerous estimated as 0.26, 0.38, 0.21, 0.12, and 0.03, respectively), resulting in a PSR closer to zero: $0.26-(0.21+0.12+0.03)=-0.11$.

\begin{figure}
\begin{center}
\includegraphics[height=5.8in,width=3.2in,angle=-90]{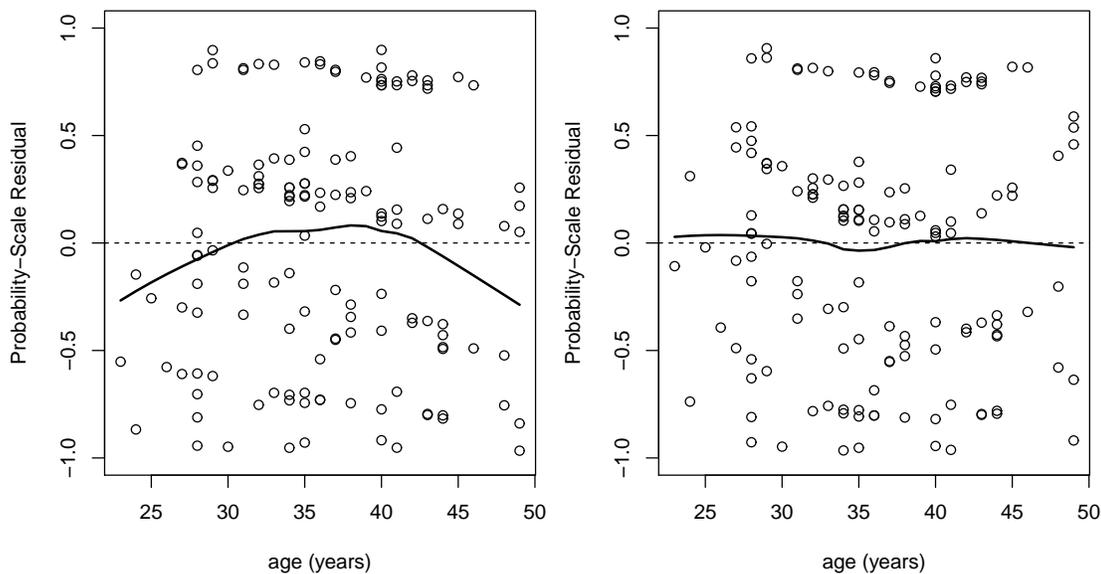}
\end{center}
\caption{Probability-scale residual-by-predictor plots with age included in a proportional odds model as a linear term (left panel) and with a linear and quadratic term (right panel).  These plots are reproduced with permission from {\it Biometrika} \citep{li2012new}.}
\label{fig1}
\end{figure}

\subsection{Biomarker Study of Metabolomics: Semiparametric Transformation Models}

HIV-positive individuals who have been on long term antiretroviral therapy (ART) appear to be at an increased risk of cardiometabolic diseases, including diabetes, compared to HIV-negative individuals.  Plasma levels of amino acids and other small molecules reflective of impaired energy metabolism, such as acylcarnitines and organic acids, were measured with mass spectrometry to provide a detailed metabolic profile for 70 non-diabetic, HIV-infected persons who were on efavirenz, tenofovir, and emtricitabine with an undetectable viral load for over 2 years \citep{koethe2016}.  There is interest in assessing associations between these biomarkers and demographic/clinical variables.  In this section, we will focus on modeling a specific biomarker, 2-hydroxybutyric acid, which is thought to be an early indicator of insulin resistance in non-diabetic persons; elevated serum 2-hydroxybutyric acid has been seen to predict worsening glucose tolerance.  2-hydroxybutyric acid is fairly skewed, ranging from 13 to 151 $\mu$M, median 34 $\mu$M in our dataset.  Even after a log-transformation, the distribution remains slightly right-skewed with some outlier levels.  Predictor variables for our model include age, sex, race, body mass index (BMI), CD4 cell count, smoking status, and ART duration (log transformed).

Because of the skewness of the biomarker outcome, we favor fitting a semiparametric transformation model, specifically $Y=T(\beta Z + \epsilon)$, where $T(\cdot)$ is an unspecified monotonic increasing transformation and $\epsilon$ is a random error with a specified parametric distribution $F_{\epsilon}$ \citep{zeng2007}.  The conditional distribution of $Y$ given $Z$ is therefore 
\begin{align*}
F_{Y|Z}(y)&=P(T(\beta Z + \epsilon) \leq y) \\
&= P( \epsilon \leq T^{-1}(y) - \beta Z)  \\
&=F_{\epsilon}(T^{-1}(y) - \beta Z).
\end{align*}  
Hence, the semiparametric transformation model can be written in a manner similar to that of the ordinal cumulative probability model, $g[F_{Y|Z}(y)]=\alpha(y)-\beta Z$, with the link function $g(\cdot)=F^{-1}_{\epsilon}(\cdot)$ and the intercept $\alpha(y)=T^{-1}(y)$.  \cite{harrell2015} has proposed using this fact to estimate parameters from the semiparametric transformation model with continuous data by maximizing an approximated multinomial likelihood, and he has implemented this procedure, denoted as orm, in R statistical software as part of his popular \lq rms' package. 

In our biomarker analysis, we fit three models of 2-hydroxybutyric acid (denoted as $Y$) on covariates: (1) a multivariable linear regression model with $Y$ untransformed; (2) a multivariable linear model with $Y$ log-transformed; and (3) a semiparametric transformation model fitted using orm with the link function $g(\cdot)$=log($-$log($\cdot))$, which corresponds to assuming $F_{\epsilon}$ follows an extreme value distribution.
Figure 2 shows quantile-quantile (QQ) plots of PSRs from each of these models compared to quantiles from a Uniform($-1,1$) distribution.  If the model is correctly specified, the residuals should be approximately uniformly distributed.  Clearly PSRs from the normal linear model are far from uniform, and although PSRs from the linear model after log-transforming the biomarker are closer to being uniform, PSRs from the flexible, semiparametric transformation model are more uniform.  

In this analysis, we could also have used OMERs to uncover lack of fit for the linear models.  However, the OMER is difficult to calculate for the semiparametric transformation model because it requires computation of the conditional expectation, and even if we went through the process of estimating the conditional expectation for all observed covariate combinations, OMERs would still be skewed and not very good for model diagnostics because the semiparametric transformation model makes no assumptions of symmetry of the OMERs, equal variance, etc.  In contrast, the PSR is easily and naturally calculated from the fitted semiparametric transformation model and makes no additional assumptions beyond that of the original model. 
Hence, the PSR is useful for comparing fit across the three different models because it is on the same scale for each; with the PSR, one is comparing apples with apples, so to speak.

\begin{figure}
\begin{center}
\includegraphics[height=6in,width=2.1in,angle=-90]{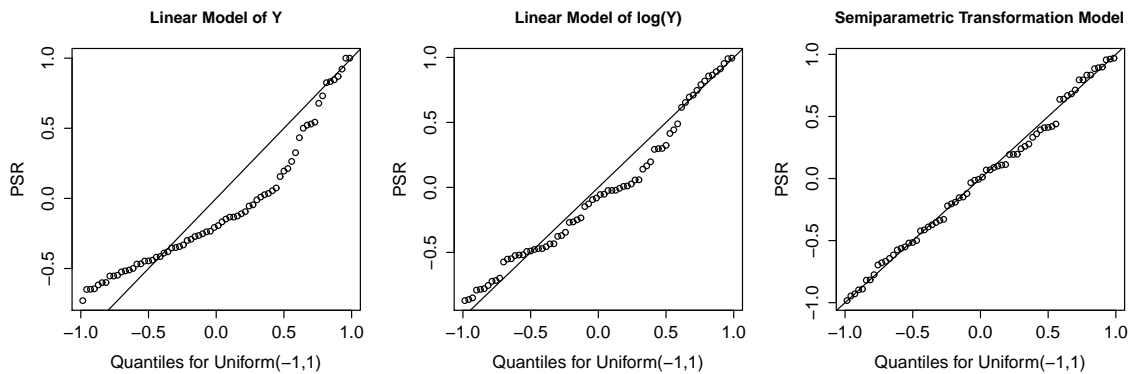}
\end{center}
\caption{QQ plots of PSRs from linear (left), linear after log-transformation (center), and semiparametric transformation models (right) of 2-hydroxybutyric acid compared to a Uniform($-1$,1) distribution.}
\label{fig2}
\end{figure}

Figure 3 shows residual-by-predictor plots for continuous covariates from the semiparametric transformation model using PSRs.  There is some evidence of non-linear relationships (top panel).  The model was re-fit expanding age, BMI, and log-transformed ART duration using restricted cubic splines with 3 knots.  Residual-by-predictor plots from these models are given in the bottom panel of Figure 3.  There is no longer evidence of non-linear residual relationships.  A likelihood ratio test confirms that the second model with the non-linear terms is a better fit (p=0.010); despite the added model complexity, the AIC for the model with the non-linear terms is lower than that without them (599 vs. 605).

\begin{figure}
\begin{center}
\includegraphics[height=6in,width=3in,angle=-90]{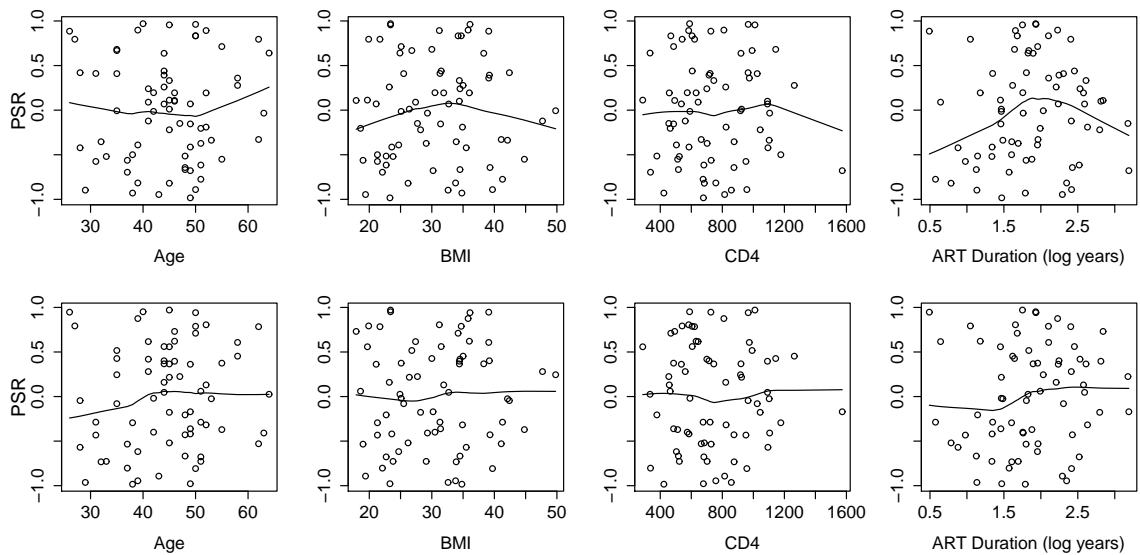}
\end{center}
\caption{Residual by predictor plots. The smoothed relationship is shown using lowess curves.  The top panels show PSRs versus continuous predictors from an initial model fit without splines.  The bottom panels show PSRs versus continuous predictors after expanding age, BMI, and log-transformed ART duration using restricted cubic splines with 3 knots.}
\label{fig3}
\end{figure}

\section{Inference}

Because the PSR is applicable to and has a common scale across a wide variety of outcome types, it can be used to test for conditional associations using residual correlation.  We describe these tests for residual correlation using PSRs in Section 4.1 and show their connection with Spearman's rank correlation in Section 4.2.  

\subsection{Tests of Residual Correlation}

We initially developed the PSR as a component of an approach for testing the association between two ordinal variables, $X$ and $Y$ while controlling for covariates $Z$ \citep{li2010test}.  Our approach was to fit separate multinomial models (e.g., proportional odds models) of $X$ on $Z$ and $Y$ on $Z$, obtain residuals from these models, and then test the association between residuals from these models.  %This is analogous to linear regression where the coefficient between $Y$ and $X$ conditional on $Z$ is equivalent to the coefficient in a regression model between OMERS from linear models of $X$ on $Z$ and of $Y$ on $Z$. 
This is analogous to linear regression where the association between $Y$ and $X$ conditional on $Z$ is captured by the correlation between OMERS from linear models of $X$ on $Z$ and of $Y$ on $Z$. We could not find a good residual for ordinal outcomes that resulted in a single value per observation and captured the necessary residual information without imposing assumptions additional to those imposed by the original model -- hence, we created the PSR.  In \cite{li2010test}, we proposed three test statistics for testing the conditional association between $X$ and $Y$, one of which was simply the sample correlation between PSRs from the two models.  We showed that our test statistics equal zero under the null hypothesis of independence between $Y$ and $X$ conditional on $Z$, and we derived their large sample distributions.  Collectively, we referred to our test statistics as COBOT (conditional ordinal by ordinal tests).  

We found that these test statistics performed well in simulations, finding a nice balance between power and robustness.  For example, if data were simulated from a proportional odds model with a linear relationship between the log-odds of $Y$ and the labels of the ordinal predictor $X$ conditional on $Z$, then our COBOT methods resulted in minimal loss of power compared to the gold standard analysis of simply fitting a proportional odds model with $X$ included as a continuous variable.  Under this scenario, the power of COBOT was much higher than models treating ordinal $X$ as a categorical variable that ignored the order information.  And when data were generated such that the log-odds of $Y$ and the labels of the ordinal predictor $X$ was non-linear conditional on $Z$, then the power of COBOT was higher than that of approaches that included $X$ in the proportional odds model as a continuous variable or a categorical variable.  Details are in \cite{li2010test}.

It would be disingenuous to claim that COBOT always outperforms other approaches.  For example, COBOT has poor power to reject the null of conditional independence when the relationship between $X$ and $Y$ conditional on $Z$ is not monotonic.  Also, subsequent simulations have suggested that the advantages of COBOT depend on the number of categories of the ordinal variable and the probability distribution of those categories.  For example, if an ordinal variable, $X$, has few categories (i.e., 2-3) then there appears to be little advantage to using COBOT over simply treating $X$ as a categorical predictor variable in a proportional odds model, whereas when there are lots of categories (i.e., $>7$), treating the ordinal variables as continuous seems to perform reasonably well, even with non-linear relationships.  These caveats noted, we believe that COBOT fills an important gap in the statistical literature regarding methods to test the conditional association between two ordinal variables while accounting for their ordinal nature.

Because of its applicability and common scale with a wide variety of outcomes, the correlation of PSRs can also be used to test for conditional associations in more general settings.  For example, using the correlation of PSRs, one could test for association between continuous, count, or ordinal $X$ and continuous, count, or ordinal $Y$ conditional on $Z$.   %Although our initial focus was to use the correlation of PSRs as a robust test statistic, it turns out that it also has a nice interpretation  as an adjusted rank correlation.

%The focus of COBOT was on ordinal outcomes and predictors.  However, given the applicability of PSRs to many different types of outcomes, similar approaches could be used test for conditional associations between more general types of variables $X$ and $Y$.  %Given the success of COBOT, we wondered whether similar approaches could be used to test the correlation between ordinal predictors and other outcome types, for example continuous outcomes.  To this end, we begin studying CoCoBOT (Conditional continuous by ordinal tests), CCountBOT (Conditional count by ordinal tests), and other types of -BOTs.  Somewhat naively, we first examined approaches such as the following:  fit a linear regression model of continuous $Y$ on $Z$, fit a proportional odds model of ordinal $X$ on $Z$, and then assess the correlation between OMERs from the first model and PSRs from the second model.  However, it quickly became apparent that the different scales of the residuals was not desirable.  Two key facts altered our thinking:  first, as demonstrated above, the PSR is well defined and easily calculated for a wide variety of variable types, and second, the connection between the correlation of PSRs and Spearman's rank correlation.

\subsection{Spearman's Partial and Conditional Rank Correlations}

Consider a model of an ordinal outcome $Y$ with a constant predictor for all subjects (i.e., a model with only an intercept).  With such a model, the predicted probability for each category is simply its empirical distribution ($n_j/n$), and the PSR is therefore a linear transformation of the ranks of $Y$.  As such, when there are no covariates, the correlation coefficient between PSRs from models for $X$ and $Y$ is equivalent to Spearman's rank correlation.  When covariates are present, the PSR can be thought of as a linear transformation of the adjusted ranks of subjects and our test statistic can be thought of as an adjusted rank correlation.  This also holds for continuous outcomes and other types of ordered discrete variables (e.g., count data), and suggests an approach for extending Spearman's rank correlation to account for covariates.  

More formally, the population parameter of Spearman's rank correlation, denoted as $\gamma_{XY}$, is the scaled difference between the probability of concordance and the probability of discordance between $(X,Y)$ and $(X_0,Y_0)$, where $X_0$ and $Y_0$ have the same marginal distributions as $X$ and $Y$, denoted $F$ and $G$, respectively, but $X_0 \perp Y_0$ and $(X_0,Y_0) \perp (X,Y)$ \citep{kruskal1958}.  With continuous $X$ and $Y$ the scaling factor is 3, which ensures that $-1 \leq \gamma_{XY} \leq 1$;  for non-continuous $X$ and/or $Y$, the scaling factor is a function of the marginal distributions of the non-continuous variables \citep{neslehova2007}.  It can be shown that this difference between concordance and discordance probabilities is equal to the covariance of PSRs (from unconditional models), and the scaling factor is simply the inverse of the square root of the product of the variances of the PSRs  \citep{liu2016}.  Specifically,
\begin{align*}
\gamma_{XY}&=c\{P[(X-X_0)(Y-Y_0)>0]-P[(X-X_0)(Y-Y_0)<0]\} \\
&=c{\text{Cov}[r(X, F), r(Y, G)]} \\
&={\text{corr}[r(X, F), r(Y, G)]}.
\end{align*}
Note, as highlighted in Section 2, that with continuous variables and correct model specification, the PSR is uniformly distributed from $-1$ to 1 and hence has variance 1/3, leading to a scaling factor of $c=3$; with discrete random variables, the variance of the PSR is $\text{Var}[r(X, F)]=(1-\sum f_x^3)/3, \text{where }f_x=P(X=x)$, which leads to the scaling factor for Spearman's correlation with discrete variables.  Therefore, Spearman's rank correlation can be written as the correlation of PSRs.  
Similarly, Spearman's rank correlation conditional on $Z$ can be defined as
\begin{align*}
 \gamma_{XY|Z}&=c_Z \{P[(X-X_0)(Y-Y_0)>0|Z]-P[(X-X_0)(Y-Y_0)<0|Z]\} \\
     &=\text{corr}[r(X, F_{X|Z}), r(Y,G_{Y|Z}) |Z],
 \end{align*}
 the conditional correlation between PSRs from models conditional on $Z$.  This expression is equivalent to Spearman's conditional rank correlation for continuous variables recently proposed by  \cite{gijbels2011}.  Unlike \cite{gijbels2011}, however, our conditional rank correlation using PSRs can also be easily applied to discrete variables.  Such a statistic describes how the rank correlation between $X$ and $Y$ varies as a function of $Z$.  
 
Finally, we define Spearman's partial rank correlation as 
 \begin{align*} 
    \gamma_{XY\cdot Z}
                  &=c^* \{P[(X-X_0)(Y-Y_0)>0|Z]-P[(X-X_0)(Y-Y_0)<0|Z]\} \\
                  &= \text{corr}[r(X, F_{X|Z}), r(Y,G_{Y|Z})],
  \end{align*}
which is a weighted average of $\gamma_{XY |Z}$.  Partial correlations describe the association between $X$ and $Y$ after adjusting for $Z$, but not as a function of $Z$.    Details are in \cite{liu2016}.

These observations fill another gap in the literature.  Pearson's partial correlation is derived as the correlation between OMERs from models of $Y$ on $Z$ and $X$ on $Z$.  When $Z$ is a single variable, this is equivalently written as ($\rho_{XY}-\rho_{XZ}\rho_{YZ})/\sqrt{(1-\rho^2_{XZ})(1-\rho_{YZ}^2)}$, where $\rho_{XY}$ denotes Pearson's correlation between $X$ and $Y$ and so forth.  The traditional Spearman's partial correlation has been proposed by substituting $\rho_{AB}$ with the corresponding rank correlations, $\gamma_{AB}$.  Although not a poor measure of association, this traditional Spearman's partial correlation is ad hoc and does not correspond with a sensible population parameter \citep{kendall1942, gripenberg1992}.  In contrast, Spearman's partial rank correlation defined as the correlation of PSRs directly corresponds to the population parameter of Spearman's rank correlation and is elegantly analogous to the definition of Pearson's partial correlation -- instead of the correlation of OMERs it is the correlation of PSRs.

The above arguments are made at the population level.  In practice, one must fit models of $Y$ on $Z$ and $X$ on $Z$ to compute the partial Spearman's rank correlation.  Given the flexibility of the PSR to a wide variety of models, the choice of models for $Y$ on $Z$ and $X$ on $Z$ are almost unlimited.  However, the choice of model is still important to ensure adequate fit for investigating residual correlation.  For example, if a model of $Y$ on $Z$ poorly fits the data, then residual correlations from this model may be misleading. To be true to the robust nature of Spearman's rank correlation, yet to be efficient, we favor fitting semiparametric models that only use the order information of the outcomes for models of $Y$ on $Z$ and $X$ on $Z$.  Specifically, we favor using the semiparametric transformation model described in Section 3.2 with $Y=T(\beta Z + \epsilon)$, where $T(\cdot)$ is an unspecified monotonic increasing transformation and $\epsilon$ is a random error with a specified parametric distribution $F_{\epsilon}$ \citep{zeng2007}.  A similar model is fit for $X$ on $Z$.  An advantage of the semiparametric transformation model is that it can be fit to binary, ordered categorical, and continuous variables.  In practice, we have used orm, introduced in Section 3.2, to obtain maximum likelihood estimates based on the semiparametric transformation model.  Extensive simulations have shown that Spearman's partial correlation using PSRs performs remarkably well with orm, even when models are misspecified (e.g., orm with a cloglog, instead of probit, link is fit to normal data) \citep{liu2016}.

Computation of conditional Spearman's rank correlations can also be computed using PSRs from semiparametric transformation models.  If $Z$ is categorical, then this correlation can simply be calculated in each level of $Z$.  With continuous $Z$, the conditional rank correlation can be computed either nonparametrically with, for example, kernal smoothers, or modeled with parametric functions \citep{liu2016}.

\subsection{Covariance of PSRs}
In COBOT, we proposed 3 test statistics.  One of them, the correlation between PSRs, has been described above as an extension of Spearman's rank correlation to account for covariates.  A second COBOT test statistic was to compare the observed values of $(X,Y)$ with the distribution of possible values under the null that $X$ and $Y$ are independent conditional on $Z$ using the difference of concordance-discordance probabilities.  %The null distribution of $(X,Y)|Z$ is the product of the distributions of $X|Z$ and $Y|Z$.  
It can be shown that this is equivalent to the expectation of the product of PSRs, $E[r(Y,F_{Y|Z}),r(X,G_{X|Z})]$, which with correctly specified models is simply the covariance of PSRs.  Hence, not surprisingly, we could also use the covariance of PSRs as a test statistic.  Although some interpretation is lost when using the covariance, it may have advantages over the correlation when modeling the rank association between $X$ and $Y$ as a function of $Z$ because it only requires modeling $E[r(X, F_{X|Z})r(Y, G_{Y|Z})|Z]$.  In contrast, models of Spearman's conditional rank correlation with discrete data also require modeling $\text{Var}[r(X, F_{X|Z})]$ and $\text{Var}[r(Y, G_{Y|Z})|Z]$.

%We proposed to If we were to compare the observed distribution with a random draw from a null distribution WORK ON THIS HERE, we are following a procedure that mimics the population parameter of Spearman's rank correlation -- only in this case our null distribution is conditional on $Z$.  [MAYBE REMOVE THE FOLLOWING: Following arguments similar to those used earlier,] it can be shown that this is equivalent to the expectation of the product of PSRs, $E[r(Y,F_{Y|Z}),r(X,G_{X|Z})]$, which with properly specified models is simply the covariance of PSRs because they have expectation 0.  Hence, not surprisingly, we could also use the covariance of PSRs as a test statistic.  Although some interpretation is lost when using the covariance, it may have advantages over the correlation when modeling the conditional correlation WORK ON THIS between $X$ and $Y$ based on $Z$ because it only requires modeling $E(X_{res}Y_{res}|Z)$, whereas models of the correlation also require modeling the variance of PSRs as a function of $Z$.

\subsection{Examples}

\subsubsection{Stage of Cervical Lesions}  

Returning to the Zambia dataset of Section 3.1, there is interest in assessing the association between stage of cervical lesions and condom use, after controlling for other variables.  Stage of cervical lesions and condom use are both ordered categorical variables.  The unadjusted Spearman's rank correlation is $-0.057$ (p-value=0.50).  Using the methods described above, Spearman's partial rank correlation was estimated to be $-0.037$ (95\% CI $-0.196, 0.123$; p-value=0.65), adjusted for age, age$^2$, CD4, education, and marital status.  

%NOT STRONG AFTER HERE:  CD4 cell count was a strong predictor of stage of cervical lesions (Spearman's partial rank correlation of XX [95\% CI XX-XX]); if CD4 cell count is removed from the model, then the partial rank correlation between stage of cervical lesions and WHO stage of HIV disease was XX (95\% CI XX-XX).  There was little evidence to suggest that the correlation between stage of cervical lesions and HIV disease differed by CD4 cell count (p=XXX).  These quantities were derived using PSRs from orm with a logit link, which with ordered categorical variables are equivalent to proportional odds models.  Confidence intervals were based on large sample approximations after using Fisher's transformation, $log[(1+\gamma_{XY\cdot Z})/(1-\gamma_{XY\cdot Z})]/2$, which typically leads to faster convergence to normality.  The test for differences in correlation based on CD4 count was performed by fitting the model $E(X_{res}Y_{res}|Z)=\alpha_0+\alpha_1 Z$ and testing the null hypothesis that $\alpha_1=0$, properly including the uncertainty in estimation of the PSRs in the computations.

\subsubsection{Biomarker Study of Metabolomics}

Returning to the biomarker study of Section 3.2, there is interest in assessing the correlation between plasma levels of various metabolites to better understand how these molecules interact among persons infected with HIV who have been on long term antiretroviral therapy.  There were 21 primary biomarkers measured on 70 HIV-infected patients.  Data were complete except for a single patient who was missing a measurement of OGIS 120.  The distributions of the biomarkers are quite heterogenous (data not shown), many are right skewed (e.g., 2-hydroxybutyric acid highlighted in Section 3.2), some have several patients with values below assay detection limits, and pairwise associations are not expected to be linear.  The biomarkers' scales vary and there is little interest in obtaining interpretable regression coefficients.  % (e.g., for a 1-unit increase in C5 acylcarnitine, hemoglobin A1C increases $x$).
For these reasons, Spearman's rank correlations between biomarkers would be ideal because of their robustness and their single number summary of the strength of association on a common scale between $-1$ and 1.  However, other variables could be associated with various biomarkers that we would like to control for, including age, sex, race, BMI, CD4 cell count, smoking status, and ART duration.  Hence, we also computed Spearman's partial rank correlation using the correlation of PSRs from models that adjusted for those variables.  This was done by fitting a model for each biomarker using a semiparametric transformation model with the covariates listed above (with ART duration log-transformed) and estimating via orm with a logit link.  (Results were very similar when using a complementary log-log link, and are not shown.)  As illustrated in Section 3.2, some of these models may have benefited from including non-linear relationships between covariates and biomarkers using splines; however, with only 70 patients, over-fitting could be an issue.  Also, these estimates are meant to be a first pass that could lead to further investigation, perhaps fine-tuning model fit using diagnostics as done in Section 3.2.  

\begin{figure}
%\begin{center}
\includegraphics[height=5.5in,width=7in,angle=0]{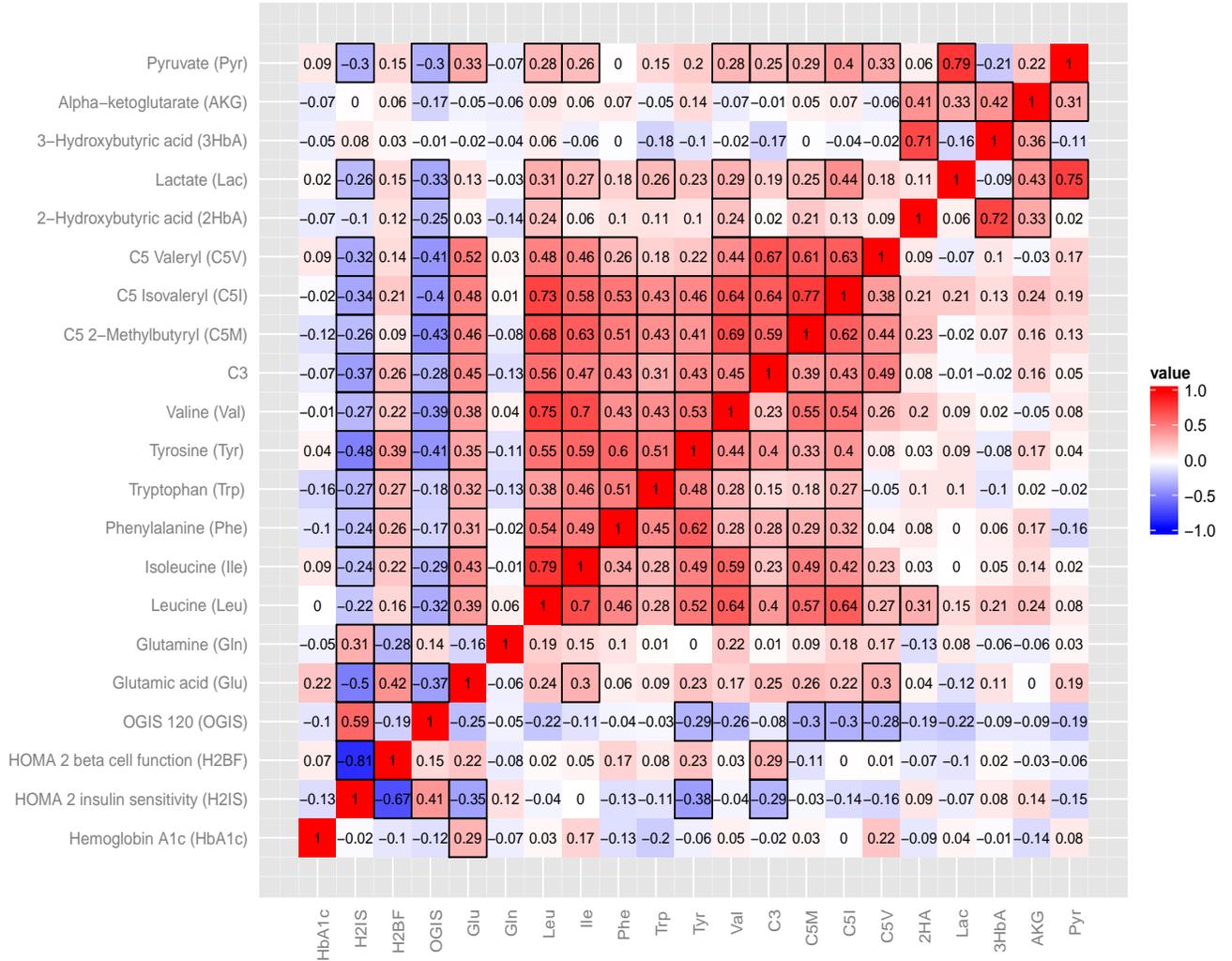}
%\end{center}
\caption{Heatmap showing the pairwise Spearman's rank correlations between 21 biomarkers.  The upper-left correlations are unadjusted, the lower-right correlations are partial correlations adjusted for age, sex, race, BMI, CD4, smoking status, and ART duration.  Colors denote the correlation with those closer to $-1$ and 1 being more blue and red, respectively.  Those correlations that are significantly different from 0 at the $\alpha=0.05$-level are boxed.}
\label{fig4}
\end{figure}

Figure 4 shows Spearman's rank correlation for all pairs of biomarkers; the quantities to the upper-left of the diagonal are unadjusted, the quantities to the lower-right of the diagonal are adjusted using the correlation of PSRs.  Red and blue shading indicates positive and negative correlation, respectively, and black boxes are drawn around those correlations that are significantly different from 0 (i.e., p-values $<0.05$).  The figure demonstrates that many of the correlations are reduced after adjusting for these additional variables.  For example, HOMA 2 insulin sensitivity appears to be correlated (positively or negatively) with many of the other biomarkers, with 15 of the 20 unadjusted pairwise correlations significantly different from 0.  However, after controlling for covariates the correlations generally weakened with point estimates closer to 0 and only 5 of 20 adjusted pairwise correlations significantly different from 0.  A similar reduction in correlation is observed for the lactate biomarker.  Although correlations generally weakened after controlling for other variables, this was not always the case.  For example, the rank correlation between hemoglobin A1C and glutamic acid increased from 0.22 to 0.29 after controlling for covariates, and the rank correlation between alpha-ketogluterate and pyruvate increased from 0.22 to 0.31 in the presence of covariates.

\subsubsection{Genome-wide Association Study}

As an additional illustration of our methods, we use the correlation of PSRs to examine the potential association between single nucleotide polymorphism (SNPs) and tenofovir clearance among patients randomized to the tenofovir/emtricitabine arm in AIDS Clinical Trials Group Protocol A5202.  Tenofovir causes kidney toxicity in some patients and there is interest in identifying SNPs that may be associated with plasma tenofovir clearance, as these SNPs may in turn be associated with risk of kidney toxicity.  An earlier genome-wide association study (GWAS) looked at the association between approximately 890,000 SNPs and tenofovir clearance using standard methods \citep{wanga2015}.  In these analyses, the association between SNPs and tenofovir clearance was modeled using linear regression adjusting for sex, age, BMI, other antiretrovirals (efavirenz or ritonavir boosted atazanavir), baseline creatinine clearance, self-reported race (white, black, Hispanic, or other), and the first two principal components for genetic ancestry.  SNPs were included in those models assuming additive effects; i.e., for bi-allelic markers with alleles A and a, genotypes A/A, A/a, and a/a were coded as 0, 1, and 2.  

Genotype can be thought of as an ordered categorical variable, and there may be benefits to treating it as such in a GWAS.  In particular, genetic effects may be additive (as assumed by the analysis model given above), dominant, or recessive, and it would be desirable to have a single analysis that is robust for detecting monotone associations between SNPs and tenofovir clearance without specifically assuming effects are either additive, dominant, or recessive.  To that end, we repeated the GWAS using the correlation of PSRs.  Specifically, we regressed tenofovir clearance on the same covariates adjusted for in \cite{wanga2015} with the exception that SNPs were not included in the model, and we also fit proportional odds models of SNPs based on these same covariates.  PSRs were then derived from all models; for the linear model we used an empirical estimate of the distribution of the residuals to compute PSRs (i.e., PSR$_i = \sum_{j=1}^n I(\hat \epsilon_j < \hat \epsilon_i)/n - \sum_{j=1}^nI(\hat \epsilon_j > \hat \epsilon_i)/n$, where $\hat \epsilon_i$ is the OMER for subject $i$).  We then computed the correlation between PSRs from the tenofovir clearance model and PSRs from the SNP models (i.e., Spearman's partial rank correlation), and computed p-values under the null hypothesis of no residual correlation.  Finally, for purpose of comparison, we repeated all analyses fitting linear models in a manner identical to that of \cite{wanga2015} except using dominant, recessive, and categorical specifications for the SNPs.  Specifically, this amounts to coding (A/A, A/a, a/a) as (0, 1, 1) for dominant, (0, 0, 1) for recessive, and using two (dummy) variables coded as (0, 1, 0) and (0, 0, 1) for categorical which therefore ignores the order information.

  %\begin{center}
  \begin{table}[ht]
  \renewcommand\thetable{1}
  \resizebox{0.95\textwidth}{!}{\begin{minipage}{\textwidth}
    \captionsetup{justification=centering}
    \caption{TDF clearance: Correlation matrix of p-values in GWAS.}     % title of Table
    \centering     % used for centering table
    \begin{tabular}{*6c}     % centered columns (6 columns)
    \hline\hline     %inserts double horizontal lines
    
    & \textbf{corr(PSRs)} & \textbf{Additive} & \textbf{Dominant} & \textbf{Recessive} & \textbf{Categorical} \\
    \hline
    %\\ [0.5ex]     % [0.5ex] adds vertical space
    % inserts table
    %heading
    \textbf{corr(PSRs)} & 1 &  &  &  &  \\     % inserting body of the table
    
    \textbf{Additive} & 0.801 & 1 &  &  &  \\
    
    \textbf{Dominant} & 0.709 & 0.734 & 1 &  &  \\
    
    \textbf{Recessive} & 0.204 & 0.326 & 0.067 & 1 &  \\
    
    \textbf{Categorical} & 0.557 & 0.649 & 0.643 & 0.653 & 1 \\
    %\\ [0.5ex]     % [0.5ex] adds vertical space
    
    \hline     %inserts single line
    \end{tabular}
    \label{table:M.comb} % is used to refer this table in the text
    \end{minipage} }
  \end{table}
  %\end{center}

Table 1 shows a pairwise Spearman's correlation matrix of the approximately 890,000 p-values using the five different analysis models. P-values between the residual correlation and additive models were strongly correlated ($\gamma=0.801$); results from the residual correlation model were also highly correlated with the dominant model ($\gamma=0.709$), less so with the categorical model ($\gamma=0.557$), and weakly correlated with the recessive model ($\gamma=0.204$).  Table 2 shows the top 10 SNPs under each analysis model, and their respective p-values. SNP rs12387850 was ranked first in the correlation of PSRs analysis (p-value $= 2.80 \times 10^{-7})$, additive (p-value $= 1.26 \times 10^{-6})$ and dominant models (p-value $= 5.63 \times 10^{-7})$, but ranked the $75^{th}$ (p-value $= 1.07 \times 10^{-5})$ and $26^{th}$ (p-value $= 3.71 \times 10^{-6})$ in the recessive and categorical models respectively. Among the SNPs significantly associated with tenofovir clearance, SNP rs12082252 ranked $1^{st}$ (p-value $= 2.28 \times 10^{-10})$ and $2^{nd}$ (p-value $= 1.69 \times 10^{-9})$ in the recessive and categorical models respectively, but was ranked much lower in the other model specifications ($346688^{th}$ using the correlation of PSRs, $1924^{th}$ in additive, $90918^{th}$ in dominant).

\begin{table}[ht]
\begin{footnotesize}
\renewcommand\thetable{2} %manual numbering of table
\captionsetup{justification=centering}
\caption{TDF clearance: SNPs with the smallest p-values in combined group analysis}
\resizebox{0.70\textwidth}{!}{\begin{minipage}{\textwidth}
\centering     % used for centering table
%\begin{tabular}{*{15}c}     % centered columns (15 columns)
\begin{tabular}{ccc|ccc|ccc|ccc|ccc}
\hline\hline     %inserts double horizontal lines
\multicolumn{3}{c|}{\textbf{corr(PSRs)}} & \multicolumn{3}{c|}{\textbf{Additive}} & \multicolumn{3}{c|}{\textbf{Dominant}} & \multicolumn{3}{c|}{\textbf{Recessive}} & \multicolumn{3}{c}{\textbf{Categorical}} \\ [0.5ex]
CHR & SNP & P & CHR & SNP & P & CHR & SNP & P & CHR & SNP & P & CHR & SNP & P  \\ [0.5ex]
\hline
2 & rs887829 & 7.08e-10 &   2 & rs887829 & 2.24e-11 &   2 & rs3755319 & 1.03e-07 &   2 & rs887829 & 4.49e-12 &   2 & rs887829 & 1.74e-12 \\ 
    2 & rs4148325 & 2.54e-09 &   2 & rs4148325 & 7.30e-11 &  19 & rs4239638 & 3.74e-07 &   2 & rs4148325 & 8.81e-12 &   2 & rs4148325 & 4.60e-12 \\ 
    2 & rs6742078 & 4.87e-09 &   2 & rs6742078 & 1.32e-10 &  19 & rs7257832 & 4.52e-07 &   2 & rs6742078 & 5.13e-11 &   2 & rs6742078 & 1.69e-11 \\ 
    2 & rs4148324 & 1.04e-08 &   2 & rs4148324 & 3.11e-10 &   2 & rs4663333 & 4.68e-07 &   2 & rs929596 & 1.02e-10 &   2 & rs929596 & 2.48e-11 \\ 
    2 & rs10179091 & 1.63e-08 &   2 & rs929596 & 3.39e-10 &   2 & rs4663967 & 5.17e-07 &   2 & rs4148324 & 1.05e-10 &   2 & rs4148324 & 4.39e-11 \\ 
    2 & rs3771341 & 3.07e-08 &   2 & rs3771341 & 3.89e-10 &   2 & rs4399719 & 8.00e-07 &   2 & rs3771341 & 3.36e-10 &   2 & rs3771341 & 6.07e-11 \\ 
    2 & rs929596 & 3.63e-08 &   2 & rs10179091 & 3.15e-09 &   2 & rs4124874 & 1.03e-06 &   2 & rs17862875 & 6.46e-09 &   2 & rs17862875 & 2.92e-09 \\ 
    2 & rs3755319 & 4.34e-08 &   2 & rs17862875 & 1.56e-08 &   2 & rs4663965 & 1.48e-06 &   2 & rs10179091 & 2.48e-08 &   2 & rs10179091 & 1.01e-08 \\ 
    2 & rs4148326 & 6.53e-08 &   2 & rs2221198 & 2.04e-08 &   2 & rs6431628 & 2.17e-06 &   2 & rs4148326 & 6.00e-08 &   2 & rs4148326 & 5.80e-08 \\ 
    2 & rs2221198 & 1.34e-07 &   2 & rs4663969 & 2.24e-08 &  11 & rs1560994 & 2.50e-06 &   2 & rs2221198 & 1.86e-07 &   2 & rs2221198 & 6.47e-08 \\ 
    2 & rs4663969 & 1.93e-07 &   2 & rs3755319 & 2.38e-08 &   2 & rs17862866 & 2.77e-06 &   2 & rs4663969 & 2.68e-07 &   2 & rs3755319 & 7.48e-08 \\ 
    2 & rs4663967 & 2.43e-07 &   2 & rs4148326 & 2.73e-08 &   2 & rs3806597 & 2.96e-06 &   2 & rs16862202 & 2.69e-07 &   2 & rs7604115 & 7.53e-08 \\ 
    2 & rs4663333 & 2.81e-07 &   2 & rs7604115 & 2.94e-08 &   2 & rs2008595 & 3.90e-06 &   2 & rs7556676 & 4.01e-07 &   2 & rs4663969 & 7.94e-08 \\ 
    2 & rs7556676 & 2.84e-07 &   2 & rs7556676 & 3.05e-08 &   2 & rs4294999 & 3.91e-06 &   2 & rs7604115 & 4.46e-07 &   2 & rs7556676 & 1.15e-07 \\ 
    2 & rs871514 & 4.07e-07 &   2 & rs871514 & 4.00e-08 &  10 & rs7915217 & 4.02e-06 &  19 & rs8111761 & 2.67e-06 &   2 & rs4663967 & 2.53e-07 \\ 
    2 & rs4294999 & 4.80e-07 &   2 & rs4663967 & 5.55e-08 &   2 & rs871514 & 4.12e-06 &   7 & rs1395381 & 2.80e-06 &   2 & rs4663333 & 2.58e-07 \\ 
    2 & rs4663965 & 5.48e-07 &   2 & rs4663333 & 5.97e-08 &   2 & rs4663963 & 4.63e-06 &   3 & rs9310867 & 4.57e-06 &   2 & rs871514 & 2.87e-07 \\ 
    2 & rs4399719 & 5.65e-07 &   2 & rs4294999 & 6.26e-08 &  19 & rs8108083 & 4.68e-06 &  12 & rs7303705 & 4.87e-06 &  14 & rs2353726 & 3.55e-07 \\ 
    2 & rs3806597 & 6.46e-07 &   2 & rs4663965 & 1.16e-07 &   6 & rs199634 & 5.13e-06 &   4 & rs3866838 & 5.14e-06 &   2 & rs4294999 & 4.28e-07 \\ 
    2 & rs7604115 & 6.46e-07 &   2 & rs4663963 & 1.33e-07 &  19 & rs2377572 & 5.64e-06 &   9 & rs7847905 & 5.53e-06 &   2 & rs4663965 & 5.99e-07 \\
\hline     %inserts single line
\end{tabular}
\label{table:combTable} % is used to refer this table in the text
\end{minipage} }
\end{footnotesize}
\end{table}

These results are fairly consistent with extensive simulations in which we generated data under different scenarios (additive, dominant, recessive, and non-linear; and with different minor allele frequencies) and investigated the power of the correlation of PSRs to detect associations under the various scenarios \citep{wanga2014}.  The correlation of PSRs had greater power to detect associations than the recessive and dominant models (except, of course, when the data were generated under these models).  The correlation of PSRs did well when the true association was additive, resulting in only a slight loss of power when compared with models that were correctly specified as additive.  However, the correlation of PSRs struggled to detect associations generated under the recessive model; categorical models were better at detecting these recessive associations and even additive models performed as well as the correlation of PSRs in the recessive case.  Hence, although it is certainly reasonable to analyze GWAS data using the correlation of PSRs, the benefits of this approach are not as apparent in this setting as in some others.  This is somewhat expected: with only three categories, the loss of power by treating ordinal SNP data as categorical is likely minimal.

\section{Discussion}

We have described a new residual, the probability-scale residual, and demonstrated its use for diagnostics and inference in several settings using actual HIV data.  We believe the PSR should be the go-to residual for the analysis of ordered categorical data and semiparametric transformation models fit to continuous data.  The PSR is also useful in many other settings that are illustrated elsewhere (see \cite{shepherd2016psr}), including time-to-event outcomes.  In some of these settings, existing residuals are available that offer similar information to the PSR.  However, as seen in Section 3.2, there are advantages to having a residual that is defined across multiple classes of models, as residuals on the same scale become easier to directly compare fit across model classes.  Because it is bounded between $-1$ and 1, the PSR is poor at detecting outliers.  (It should be noted that in some analyses [e.g., semiparametric transformation models and quantile regression] the analysis model down weights the influence of outliers, so residual plots that are dominated by \lq outliers' are actually inconsistent with the model that was fit and may not be desirable.)  If outlier detection is desired, a simple transformation of the PSRs (e.g., $\Phi^{-1}\{(PSR+1)/2\}$ with $\Phi^{-1}(\cdot)$ denoting the inverse cdf of the standard normal distribution) allows their detection.  There are certainly other diagnostics for which residuals more specialized than the PSR might be more appropriate.

The extension of Spearman's rank correlation to adjust for covariates using the correlation of PSRs is an important development.  In many settings, researchers are not interested in regression coefficients, but would like a simple, single number summary of the strength of relationship between variables, after controlling for other variables, that is given in a constant scale regardless of the type of outcome.  Our definition of Spearman's partial rank correlation provides such a summary measure, and it estimates a sensible population parameter.

The examples described in this chapter have focused on univariate outcomes.  Of particular interest would be whether some of these approaches could be of value with multivariate data, for example in the analysis of repeated measures data.  We are also interested in examining the utility of Spearman's partial rank correlation using PSRs where at least one of the outcomes is a time-to-event.  These represent areas of future research.

\section{Acknowledgements}

We would like to thank Vikrant Sahasrabuddhe, John Koethe, and the AIDS Clinical Trials Group for giving us permission to use their data.  This work was funded in part by the National Institutes of Health, grant numbers R01 AI93234 and P30 AI110527.

\bibliographystyle{apa}
\bibliography{chapter-ref}

\end{document}